\documentclass{emulateapj}
\usepackage{epsfig}
\usepackage{longtable}

\begin{document}

\title{How Massive are Massive Compact Galaxies?}

\author{Adam Muzzin\altaffilmark{1}, Pieter van
  Dokkum\altaffilmark{1}, Marijn Franx\altaffilmark{2}, Danilo
  Marchesini\altaffilmark{1}\altaffilmark{3}, Mariska
  Kriek\altaffilmark{4} \& Ivo Labb\'{e}\altaffilmark{5}}

\altaffiltext{1}{Department of Astronomy, Yale
  University, New Haven, CT, 06520-8101; adam.muzzin@yale.edu} 
\altaffiltext{2}{Leiden Observatory, Leiden University, PO Box 9513,
  2300 RA Leiden, Netherlands} 
\altaffiltext{3}{Department of Physics and Astronomy, Tufts University, Medford, MA 02155} 
\altaffiltext{4}{Department of Astrophysical Sciences, Princeton
  University, Princeton, NJ, 08544} 
\altaffiltext{5}{Hubble Fellow, Carnegie Observatories, 813 Santa
  Barbara Street, Pasadena, CA, 91101} 
\begin{abstract}
Using a sample of nine  massive compact galaxies at $z \sim$ 2.3 with
rest-frame optical spectroscopy and 
comprehensive U $\rightarrow$ 8$\micron$ photometry we 
investigate how assumptions in SED modeling change
the stellar mass estimates of these galaxies, and how this affects our 
interpretation of their size evolution.  The SEDs are fit to
$\tau$-models with a range of metallicities, dust laws, as well as 
different stellar population synthesis codes.  
These models indicate masses equal to, or slightly smaller than our 
default masses. The maximum difference is 0.16 dex for each parameter
considered, and only  0.18 dex for the most extreme combination of parameters.  
Two-component populations with a maximally old stellar population superposed with a young
component provide reasonable fits to these SEDs
using the models of Bruzual \& Charlot (2003); however, using models with
updated treatment of TP-AGB stars the fits are poorer.  The two-component models predict masses that are
0.08 to 0.22 dex larger than the $\tau$-models.  We also test the effect of
a bottom-light IMF and find that it would reduce the
masses of these galaxies by 0.3 dex.  Considering the range of allowable
masses from the $\tau$-models, two-component fits, and IMF, we
conclude that on average these galaxies lie below the mass-size
relation of galaxies in the local universe by a factor of 3-9, depending on the SED models used. 
\end{abstract}


\section{Introduction}
Numerous observational studies have shown that the population of
massive galaxies at $z \sim$ 1.5 - 2.5 is significantly more compact
than local galaxies of similar mass (e.g., Daddi et al. 2005; Trujillo
et al. 2006, Zirm et al. 2007; Toft et al. 2007, Longhetti et
al. 2007, Cimatti et al. 2008, Damjanov et al. 2008, van Dokkum et
al. 2008, Franx et al. 2008, Buitrago et al. 2008, van der Wel et
al. 2008, Saracco et
al. 2009, Taylor et al. 2009).  Simple arguments based on the light profiles of these compact
galaxies (e.g., Bezanson et al. 2009, Hopkins et al. 2009), as well as
the evolution of their space density (e.g., van der Wel et al. 2009,
Saracco et al. 2009) suggest that these galaxies may be the
progenitors of local early-type galaxies being assembled from
the inside out, and these arguments are supported by recent numerical
simulations (e.g., Naab et al. 2009).
\newline\indent
Thus far, little effort has been focused on how well-determined the stellar
masses (M$_{\mbox{\scriptsize star}}$) of these galaxies are.  Indeed, claims about the
total size growth of the galaxies, as well as
explanations for the method of size growth based on their space density require us to
associate these galaxies to local galaxies of a particular mass. 
It is well known that there can be serious systematic effects in
photometrically-determined  M$_{\mbox{\scriptsize star}}$'s caused by assumptions about
metallicity, the galactic extinction law, or the method of stellar population
synthesis (e.g., Maraston et al. 2006, Conroy et al. 2008a; 2008b, Marchesini et al. 2009,
Longhetti \& Saracco 2009, Muzzin et al. 2009).  Furthermore, most models of high-z galaxies
assume only simple star formation histories (SFH; usually
parameterized by an exponentially-decreasing SFR with timescale 
$\tau$).  As suggested by earlier authors (e.g., van Dokkum et
al. 2008, Hopkins et al. 2009, La Barbera \& De Carvalho 2009), if the compact galaxies have
multi-age stellar populations with different spatial distributions, this may complicate our
interpretation of their evolution.  In this letter we investigate how robust the M$_{\mbox{\scriptsize star}}$ of
$z \sim$ 2.3 compact galaxies are, and how the uncertainties on these
masses affect our interpretation of their size evolution.
\begin{figure*}
\plotone{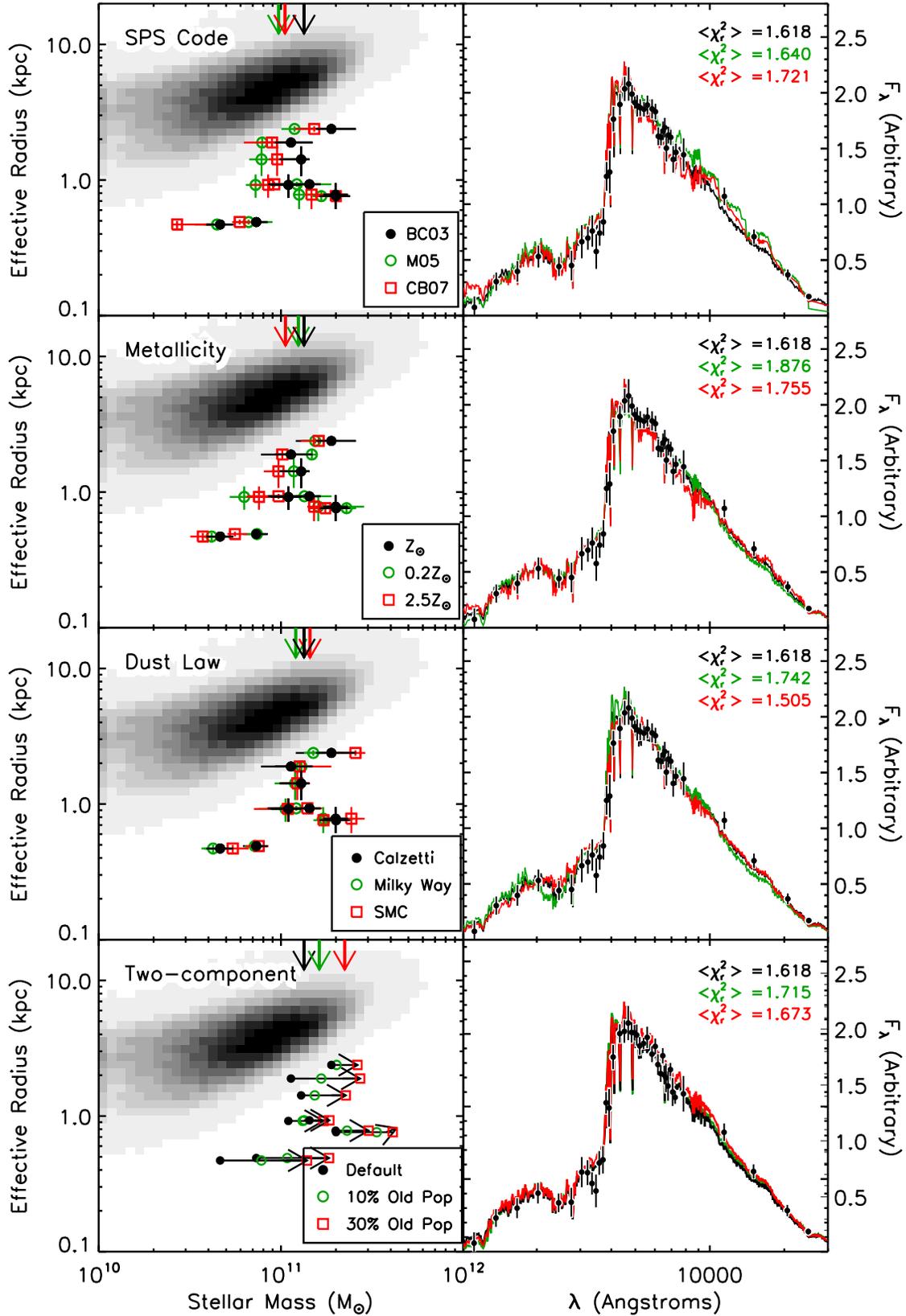}
\caption{\footnotesize Left panels: Plots of effective radius
  (R$_{e}$) vs. stellar mass (M$_{\mbox{\scriptsize star}}$) for various assumptions in
  the SED modeling.  Local galaxies from the SDSS are plotted as
  contours ranging between 5 galaxies (lightest) to 1000 galaxies (darkest).  The default model for the compact galaxies are plotted as
  large black circles, with variations denoted in the legend.  The
  average mass for each model is shown as an arrow at the top of the
  panel.  The
  effects of SED modeling assumptions on the M$_{\mbox{\scriptsize star}}$ of the
  galaxies is less than 0.30 dex for all tested assumptions
  showing that size growth at fixed mass of a factor of 3-9 is a robust result from
  these data.  Right Panels: Mean SEDs for the galaxies with the best
  fits overplotted.  The  $\langle\chi_{r}^2\rangle$
  are similar for all fits, which shows that  assumptions in the
  SED modeling such as SPS code, metallicity, dust law, or
  two-components cannot be fit as free parameters, even with these
  well-constrained SEDs.}
\end{figure*}
\section{Data \& SED Fitting Method}
Our sample of galaxies consists of the nine compact galaxies with
strongly suppressed star formation presented in Kriek et al. (2006; 2008).
The relatively low star
formation rates of these galaxies have been recently reconfirmed both
by ultradeep spectroscopy (Kriek et al. 2009), and SED 
modeling that includes the rest-frame NIR (Muzzin et al. 2009).  These galaxies
have effective radii (R$_{e}$) of $\sim$ 1 kpc as measured from 
NICMOS H$_{F160W}$-band imaging by van Dokkum et al. (2008).  The
combination of accurate R$_{e}$'s from space-based imaging, as well as the spectroscopic redshifts
and well-sampled SEDs make this the best sample for testing the 
effect of assumptions in SED modeling of M$_{\mbox{\scriptsize
    star}}$ on the inferred size evolution of massive galaxies.  
\newline\indent
For these galaxies we adopt two methods for fitting the SEDs.  In
order to test the effects of metallicity, dust law, and SPS code on
the M$_{\mbox{\scriptsize star}}$ we fit the galaxies to models with exponentially
declining SFHs, hereafter ``$\tau$-models''.  In these models $\tau$, age,
and A$_{v}$ are fit as free parameters.  The SED fitting is performed using a
$\langle \chi^{2}\rangle$-minimization routine and the errors in M$_{\mbox{\scriptsize
    star}}$ are determined using Monte Carlo simulations.  Details of the SED fitting procedure and
parameter error estimation are described in Muzzin et al. (2009).  In attempt to isolate variables
we adopt the models of Bruzual \& Charlot (2003; BC03) with solar
metallicity, the Calzetti et al. (2000) dust law, and a Chabrier
(2003) IMF as the control
model.  This combination of models and parameters has been used
extensively in previous studies of compact galaxies, and was the default model for
the van Dokkum et al. (2008) study, making it the obvious choice for a
control model.  
We compare the M$_{\mbox{\scriptsize star}}$ from the control
model to those determined with the dust law from the SMC (Pr\'{e}vot
et al. 1984) and Milky Way (Fitzpatrick et al. 1986), a subsolar
(0.2Z$_{\odot}$) and supersolar (2.5Z$_{\odot}$) metallicity, as well as M$_{\mbox{\scriptsize star}}$ from the SPS codes of Maraston (2005; hereafter M05), and
Charlot \& Bruzual (in preparation, hereafter CB07).
\newline\indent
We also test the effects of a two-component stellar population on the M$_{\mbox{\scriptsize star}}$ of the
galaxies.  
The SEDs are fit with a linear combination of a ``young'' and ``old''
component.  The young component has $\tau$,
age, and A$_{v}$ as free parameters, but age restricted to $<$ 0.5
Gyr. The old component is a maximally old stellar population with no
dust, and therefore has no free parameters.  We perform the fitting three times, varying the amplitude of the
old component between 10, 20, or 30\% of the total observed H$_{F160W}$-band light.  We note this does not define the
full range of M$_{\mbox{\scriptsize star}}$ that could be allowed from two-component models; however, the primary purpose of fitting these models is  to explore
the possibility of components with very different M/L ratios that
could plausibly explain the sizes of the compact galaxies using age gradients.
\section{Effects of Assumptions in SED Modeling}
In the left panels of Figure 1 we show the effects of the different assumptions in the
SED modeling on the M$_{\mbox{\scriptsize star}}$ of the compact galaxies by comparing their location in
the R$_{e}$ vs. M$_{\mbox{\scriptsize star}}$ plane to galaxies from the SDSS
(adapted from Kauffman et al. 2003; Franx et al. 2008).  In the right
panels we show the mean observed SED of the entire sample as well as
the mean fit.  The $\langle \chi_{r}^{2}\rangle$ of all 
fits are similar demonstrating that even with these high quality SEDs it is
not possible to fit SPS code, metallicity, dust law, or multiple
components as free parameters.  
\newline\indent
Compared to the BC03 control model, the masses from the SPS codes of
M05 and CB07 are systematically lower by
0.13 dex.   Assuming a subsolar or supersolar metallicity systematically
reduces the $\langle$M$_{\mbox{\scriptsize star}} \rangle$ by 0.10 and 0.16 dex compared to their control values, respectively.  The Milky Way dust law
reduces the $\langle$M$_{\mbox{\scriptsize star}} \rangle$ by 0.03 dex, whereas the SMC dust law
increases it by 0.06 dex.  
Figure 1 shows that although all the SED modeling assumptions cause
systematic differences in the M$_{\mbox{\scriptsize star}}$ of the galaxies, none of them are large enough
to change the fact that the galaxies are significantly more compact
than local galaxies of similar mass.  
In fact, even if we combine the most extreme assumptions
in the SED modeling, e.g., the M05 models with supersolar metallicity,
and the Milky Way dust law, the mean M$_{\mbox{\scriptsize star}}$ is
only 0.18 dex lower than the default modeling assumptions.  
\newline\indent
The bottom panel of Figure 1 shows the effect of the two-component
models on M$_{\mbox{\scriptsize star}}$.  Adding an old stellar
component to the modeling increases the overall
M$_{\mbox{\scriptsize star}}$ by 0.08 to 0.22 dex.  Interestingly,
two-component fits from the models of BC03 have a better $\langle
\chi^2_{r} \rangle$ than from the M05 and CB07 models.  In the left panel of Figure 2 we plot
two-component fits using all three SPS codes.  In the right panel we
plot the average old and young components of the
fits using the BC03 models.  Two-component fits
from M05 and CB07 models are probably poorer descriptions of the data because of the
increased contributions from the TP-AGB stars in those models.  These
stars contribute significantly to the rest-frame NIR flux in young populations
(0.2 - 2 Gyr), and therefore, in these models the SED of both the young and old
component have significant flux in the rest-frame NIR, and their
linear combination cannot reproduce the
overall SED shape.  
This suggests that models that explain the dramatic size evolution of these galaxies
using age gradients in local galaxies (e.g., La Barbera \& De
Carvalho 2009) are unlikely to be compatible with our observed SEDs if
interpreted with the M05 and CB07 models.
\begin{figure*}
\plotone{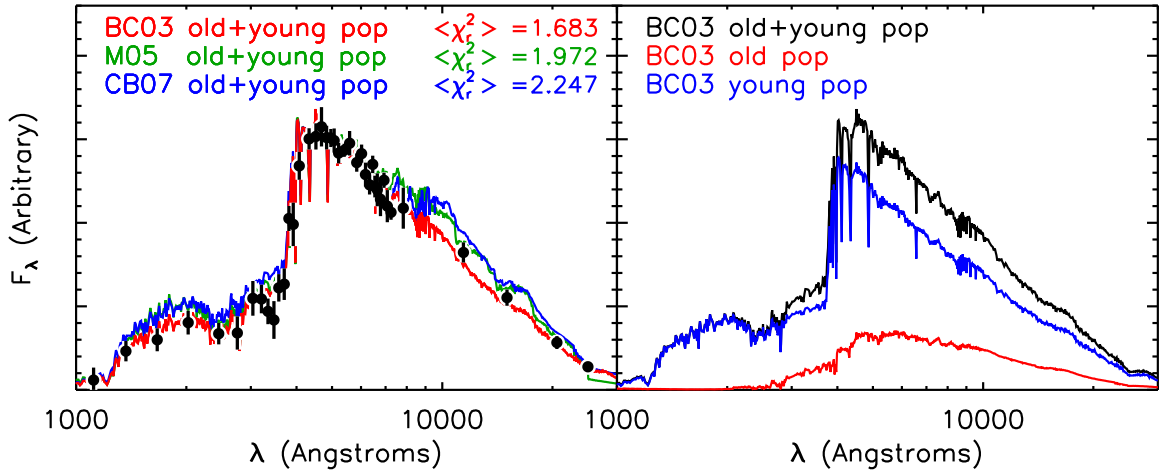}
\caption{\footnotesize Left Panel: Mean SED for the compact galaxies
  with the best-fit two-component model from the various SPS codes
  superposed.  The two-component model has 20\% of the observed
  H$_{F160W}$-band light in an old component.  The BC03 model provides
  a good fit to the data, but the fits with the M05 and CB07
  models are poorer, suggesting that they are less compatible with a
  two-component stellar population for these galaxies.  Right Panel:
  Breakdown of the BC03 SED in the left panel into the old and young
  components.}
\end{figure*}
\section{Implications for the Size Growth of Massive Galaxies}
For the single-component $\tau$-models, it is clear that changes 
in the SED modeling assumptions produce M$_{\mbox{\scriptsize star}}$ values that are
typically smaller on average than the default  model.  In particular, if we believe
the newer SPS models of M05 and CB07 provide more accurate masses than
the BC03 models, then it has two implications for the method of size growth for these galaxies.  
\newline\indent
First, assuming our
sample of galaxies increase in M$_{\mbox{\scriptsize star}}$ by a
factor of $\gtrsim$ 2 from $z \sim$ 2.3 to $z =$ 0, then to match the sizes of
local galaxies they would have to
increase in size by a factor of $\gtrsim$ 5 using the default model,
and a factor of $\gtrsim$ 4  for the
newer SPS models.  Recent
numerical simulations show size growth of a factor 
of $\sim$ 3, with an M$_{\mbox{\scriptsize star}}$ increase of a
factor of $\sim$ 2 since $z$ = 2 (e.g., Naab et al. 2009), and
therefore are in modestly better agreement with the size-mass relation from the
newer SPS models.  Second, the lower M$_{\mbox{\scriptsize star}}$ from the
M05 and CB07 models also allow for more
simultaneous mass and size growth in the
galaxies.  Bezanson et al. (2009) argued that given the M$_{\mbox{\scriptsize star}}$ and
number density of the $z \sim$ 2.3 galaxies, minor mergers are the
best candidate for increasing the size of these galaxies because they
increase the size faster than the M$_{\mbox{\scriptsize star}}$.  If
the compact galaxies have an $\langle$M$_{\mbox{\scriptsize star}} \rangle$
that is 0.13 dex lower, the mass and number density 
argument still favors minor merging as the best candidate for
size growth; however, it would permit some fraction of the population to
grow in size from major mergers.  
\newline\indent
Recently, Mancini et al. (2009) claimed that using the newer SPS
models would reduce the offset of our galaxies from the local relation
by a factor of $\sim$ 3, and inferred that this could significantly alter our interpretation of
their size evolution.  Our modeling shows that this offset
is only a factor of $\sim$ 1.2, and hence does not significantly alter
the compact galaxy ``problem''.  Mancini et al. (2009) also claimed that with the newer masses
and a 25\% systematic underestimate in sizes that some of the galaxies
in our sample may follow the local relation; however, Figure 1 
shows that this is not the case, and that even the galaxies that lie closest to
the local relation are still a factor of $\gtrsim$ 2 off the relation.
\newline\indent
If the two-component models provide the correct masses, they suggest very
different scenarios for size growth.  Assuming the spatial distribution of the old and young 
components is identical, then the size growth at fixed mass
would be significantly larger than previous estimates, a factor of $\sim$ 9.
Furthermore, these galaxies would be roughly as massive as the most
massive galaxies in the local sample, which means that the size growth
would have to be extremely efficient, resembling a pure
expansion model.   
\newline\indent
Such efficient size growth may not be required if the old
component is significantly more extended than the young component.  If
so, the mass-weighted size of the galaxies could be much larger than its light-weighted size.
In our SED fits the old component of these galaxies contain 10-30\% of the total light
(by construction), yet it contains 50-80\% of the total stellar mass.
Indeed, we cannot rule out the possibility that if
the old component has an R$_{e}$ of $\sim$ 10 kpc and 50-80\% of the
mass, that we may be looking at fully formed early-type galaxies
with compact, centrally concentrated poststarburst component in the
center.  However, 
local early-type galaxies tend to have only modest color gradients, most of which can
be attributed to metallicity (see e.g., Franx \& Illingworth 1990, Sanchez-Blazquez et al. 2007,
Kormendy et al. 2009). These color gradients imply that galaxies have
decreasing mass-to-light ratios with radius, and are therefore more
compact in mass than in light, not the converse.  It is possible that
a combination of mergers and low-level star formation could
mix the composite populations and create the locally-observed color gradients; however, with the
two-component SED modeling the $z \sim$ 2.3 galaxies are already as massive
as the most massive local galaxies, leaving little room for structural
changes caused by processes that require additional mass growth.
\section{IMF or AGN?}
Another effect that could account for the extreme
R$_{e}$-M$_{\mbox{\scriptsize star}}$ relation of the distant galaxies is an IMF that becomes increasingly bottom-light at
higher redshift.  Such an IMF has been suggested for high redshift
early-type
galaxies (van Dokkum 2008).  If we fit
our galaxies using the M05 models and the van Dokkum (2008) bottom-light IMF
we find that the $\langle$M$_{\mbox{\scriptsize star}} \rangle$ is lower by
0.30 dex.  
However, even with masses this low, these galaxies would still be a
factor of $\sim$ 3 smaller in size than local galaxies of similar
mass.  
\newline\indent
It is also worth noting that none of the compact galaxies show an observed 8$\micron$ excess (rest-frame
$\sim$3$\micron$) above the best-fit SED template.  These excesses are not uncommon in
high redshift galaxies (e.g., Labb\'{e} et al. 2005; Donley et
al. 2008, Mentuch et al. 2009, Muzzin et al. 2009) and could be
caused by an AGN.
Combined with the fact that none of the galaxies show emission lines,
it suggests that the size measurements of these galaxies are not contaminated by an unresolved point source (see also Kriek et al. 2009).\section{Discussion}
Overall, our modeling shows that despite the numerous systematic
uncertainties involved in determining photometric M$_{\mbox{\scriptsize star}}$'s, the
most extreme masses of the $z \sim$ 2.3 compact galaxies are 0.18 dex different from our nominal
estimates using a normal IMF, and 0.30 dex different using a bottom-light IMF.  From this, we conclude that on average these galaxies lie a
factor of $\sim$ 3 - 9 below this size-mass relation of local
galaxies, depending on the SED models used.  This range could be
slightly larger because our comparisons have been made to
the default SDSS model.  Although using different SPS models is unlikely to affect
the M$_{\mbox{\scriptsize star}}$ of the SDSS galaxies because at old
ages the BC03 and M05 models produce similar masses (e.g., M05), and the
dust content of the massive SDSS galaxies should be negligible, there
could be small systematic offsets between the SDSS M$_{\mbox{\scriptsize star}}$
and our M$_{\mbox{\scriptsize star}}$
because of technical issues such as fiber aperture
corrections, or the method of measuring total magnitudes. 
We note that other systematic effects may play a role. In particular,
there may be systematic differences in the measurement of effective
radii between the high and low redshift samples. van Dokkum et al.\
(2008), van der Wel et al.\
(2008), and Hopkins et al.\ (2009) show that biases due to surface
brightness effects are probably small, but deeper data at high
redshift would obviously be helpful. The low redshift data also
have significant uncertainties: interestingly, Guo et al.\
(2009) find that the sizes of the most luminous galaxies in the
SDSS NYU-VAGC (Blanton et al.\ 2005) are likely underestimated,
which would imply somewhat stronger evolution than the factor 3-9
that we find here.
\newline\indent
With our data we still cannot rule out age (and hence M/L) gradients in these
galaxies, which could bias the size measurements of these galaxies;
however, such two-component models are disfavored by our data using the most recent SPS
codes.  A two-component model is still compatible with our SEDs when
using the BC03 models, therefore measuring color gradients in these
galaxies out to large radii will be valuable for understanding if the
luminosity-weighted sizes of these galaxies are comparable to their mass-weighted sizes.  
\newline\indent
Although the range of allowed photometric-M$_{\mbox{\scriptsize
    star}}$ is only a few tenths of a dex, it is
large enough that we advise caution when using the number density
of galaxies larger than a given mass to identify the descendants of these galaxies at
various redshifts.  These galaxies sit on the exponential tail of
the mass function (e.g., Marchesini et
al. 2009) and therefore small, systematic differences in their
photometric-M$_{\mbox{\scriptsize star}}$ like those shown here could result in
significant errors in their space density.  
Ultimately, dynamical measurements are needed to calibrate photometric-M$_{\mbox{\scriptsize
    star}}$'s at high redshift.  Early results suggest the agreement
    between the two is reasonable (e.g., van Dokkum et al. 2009,
    Cappellari et al. 2009); however, it will be very important to
    extend these to large samples with a range of properties.
\acknowledgements
We thank Claudia Maraston for providing the bottom-light IMF model, and S. Charlot for providing the unpublished CB07 stellar
population synthesis models.  AM gratefully acknowledges financial support for this work
from an NSERC PDF fellowship.  The authors acknowledge support from
NSF CAREER AST9449678, and Spitzer/JPL grants RSA 1277255, 1282692,
and 1288440.

\end{document}